\documentclass[10pt,twocolumn]{IEEEtran} 

\ifCLASSINFOpdf
\else

\usepackage[dvips]{graphicx}
\fi
\usepackage{url}
\usepackage{graphicx}
\usepackage{diagbox}
\usepackage{balance}
\usepackage{multirow, cite}
\usepackage{xcolor,colortbl, soul}
\usepackage{tabularx,booktabs}
\usepackage[utf8]{inputenc} 
\usepackage[T1]{fontenc}
\usepackage{url}
\usepackage{ifthen}
\usepackage{cite}
\usepackage[cmex10]{amsmath} 
\usepackage{subfigure}
\usepackage{cite}
\usepackage{amsmath,amssymb,graphicx,epsfig,fancyhdr,amsthm,tabulary,epstopdf}
\usepackage{algorithm}
\usepackage{algorithmic}
\usepackage{pdfsync}
\usepackage[justification=centering]{caption}
\theoremstyle{remark}

\theoremstyle{definition}

\newcommand{\CASE}[1]{\STATE \textbf{case} #1\textbf{:} \begin{ALC@g}}
	\newcommand{\ENDCASE}{\end{ALC@g}}

\newcommand{\DEFAULT}{\STATE \textbf{default:} \begin{ALC@g}}
	\newcommand{\ENDDEFAULT}{\end{ALC@g}}
\newcommand{\DEFAULTLINE}[1]{\STATE \textbf{default:} }

\newcommand\Tstrut{\rule{0pt}{2.6ex}}	

\begin{document}
\bstctlcite{IEEEexample:BSTcontrol}
\title{Secure Precoding in MIMO-NOMA: \\ A Deep Learning Approach}

\author{Jordan Pauls, \IEEEmembership{Student Member, IEEE}, and Mojtaba Vaezi, \IEEEmembership{Senior Member, IEEE} 
}

\author{Jordan~Pauls,~\IEEEmembership{Student~Member,~IEEE,}~and~Mojtaba~Vaezi,~\IEEEmembership{Senior~Member,~IEEE} 
	\thanks{The authors are with the Department
		of Electrical and Computer Engineering, Villanova University, Villanova,
		PA 19085 USA (e-mail: jpauls1@villanova.edu; mvaezi@villanova.edu).}
}

\maketitle

\begin{abstract}
A novel signaling design for secure transmission over two-user multiple-input multiple-output  non-orthogonal multiple access channel using deep neural networks (DNNs) is proposed. The goal of the DNN is to  form the covariance matrix of users' signals  such that  the message of each user is transmitted reliably while being confidential from its counterpart.
The proposed DNN linearly {precodes each user's signal} before superimposing them and 
achieves near-optimal performance with  significantly lower run time. 
 Simulation results show that the proposed models reach about 98\% of the secrecy capacity rates. The spectral efficiency of the DNN precoder is much higher than that of existing analytical linear precoders--e.g., generalized singular value 
 decomposition--and its on-the-fly complexity is several times less than the existing iterative methods.

\end{abstract}


\IEEEpeerreviewmaketitle

\section{Introduction}

Non-orthogonal multiple access (NOMA) is  a promising candidate for connecting massively increasing devices to fifth-generation  and beyond wireless networks \cite{vaezi2019multiple}.
 NOMA is the optimal transmission strategy for both the single-input, single-output (SISO) and multiple-input, multiple-output (MIMO) cases in a  single-cell network.
 In the SISO case, superposition coding at the transmitter with successive interference cancellation at the receiver is optimal. In contrast, in the MIMO case, \textit{dirty-paper coding} (DPC) is the optimal solution.  Nonetheless, in both cases, the base station (BS)  \textit{broadcasts} a superimposed signal of multiple users.  This makes secure communications challenging in the presence of adversarial users as signals can be eavesdropped on by such users.   
 
 \textit{Physical layer security}  
 enables the exchange of confidential messages over a wireless
 medium in the presence of internal or unauthorized eavesdroppers  \cite{mukherjee2014principles}.
Specifically, in two-user MIMO-NOMA networks, both users can transmit their messages  concurrently and confidentially via \textit{secret dirty-paper coding} (S-DPC) 
 	\cite{ekrem2012capacity}. 
  While S-DPC is the most spectral efficient precoding for the  two-user MIMO-NOMA, it is excessively complex for practical uses.
Alternatively, S-DPC region can be achieved by \textit{linear precoding} and power allocation schemes.   
  In the past years, various linear precoding schemes have been introduced  \cite{fakoorian2013optimality,park2015weighted,qi2020secure,hanif2019robust}. \textit{Generalized singular value 
  	decomposition} (GSVD)-based  precoder \cite{fakoorian2013optimality} is a fast analytical precoder, but it falls short of getting the capacity region when the users have a single antenna. Also, weighted sum-rate maximization
 \cite{park2015weighted} and power-splitting  \cite{qi2020secure} approaches are still too time-consuming to be used in practice. More accurately, these approaches require much higher time than the \textit{coherence time} of wireless channels which is about a few milliseconds \cite{tse2005fundamentals}.

This letter  exploits deep learning (DL) to design 
the covariance matrices of the channel input vectors--or equivalently, {to design precoding and power allocation matrices--for secure  MIMO-NOMA transmission.}    Embedding DL into the mobile and wireless networks is well justified in various cases, e.g., when closed-form solutions require poor approximation or the complexity of
existing techniques is high \cite{zhang2019deep,zhang2019deepwiretap,gumus21finite}. 

\begin{figure}[t]
    \centering
	\includegraphics[width=0.35\textwidth]{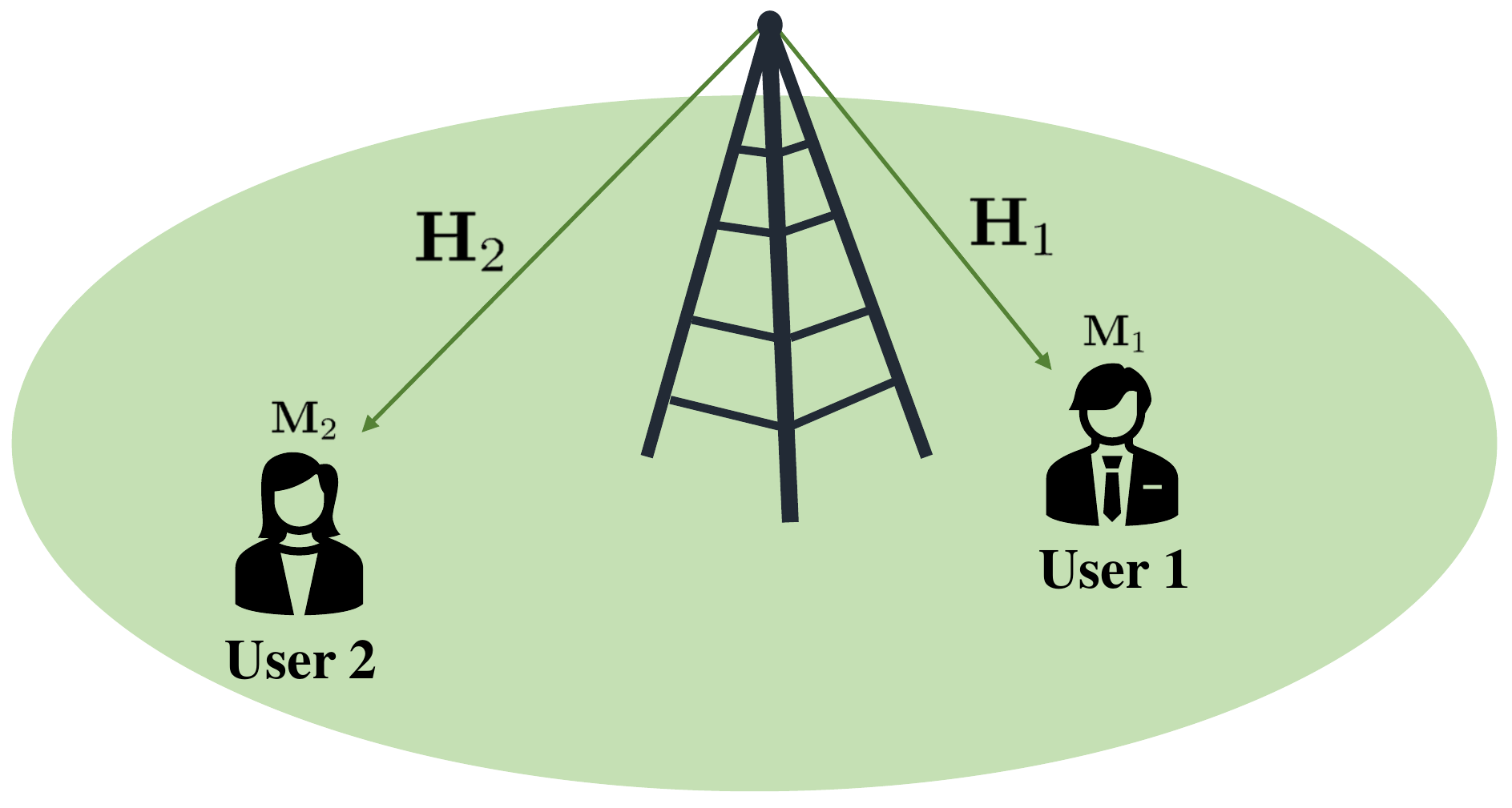}
	\caption{Two-user MIMO-NOMA with confidential messages.  $M_1$ and $M_2$ are the messages for user~$1$ and user~$2$, respectively. Each user is intended to decode its own message but not the other one. }
	\label{fig:model}
\end{figure}

We leverage supervised deep neural networks  (DNNs) for secure communication for the two-user MIMO-NOMA channel (see Fig.~\ref{fig:model}), resulting in a {significantly faster solution while almost reaching the spectral efficiency of the S-DPC.} 
To fulfill the task, we  first decompose the two-user MIMO-NOMA with confidential messages into two wiretap channels \cite{qi2020secure} and use a wiretap channel solution for generating and labeling the training data set. 
We then build and train DNN models that learn to approximate the function mapping channel matrices, 
 base station power, and required secrecy rate of the users  {to the covariance matrix  of the channel input vector for each user.}  
 Simulation results prove the efficacy of the developed model since the proposed DL-based precoding scheme has near-optimal performance and outperforms GSVD-based precoding with a large margin. It also brings mapping time below the coherence time of the wireless channel. Therefore, the channel input signal can be designed before the channel coefficients become stale, which is crucial in practice.

\section{System Model and Existing Solutions}

Consider  a single-cell  two-user MIMO-NOMA network, as shown in 
Fig.~\ref{fig:model}. Assume that the transmitter, user~$1$, and user~$2$  are equipped with $n_t$, $n_1$, and  $n_2$ antennas, respectively. The transmitter wishes to send two messages $M_1$ and $M_2$ to user~$1$, and user~$2$, respectively. 
When PHY security is a concern,  
each message must be kept \textit{confidential} from the other user  \cite{liu2010multiple}.
\footnote{In information theory, this channel is known as the MIMO broadcast channel (BC) with two 
	confidential messages \cite{liu2010multiple}.} That is, user~$1$ should not be able to decode $M_2$ and vice versa. 
  As such, the  transmitter \textit{securely} encodes $M_1$ and $M_2$ to codewords $\mathbf{x}_1 \in \mathbb{R}^{n_t 
	\times 1}$ and $\mathbf{x}_2 \in \mathbb{R}^{n_t 
	\times 1}$, \textit{superimposes} them $\mathbf{x} = \mathbf{x}_1 +\mathbf{x}_2 $, and \textit{broadcasts} $\mathbf{x}$ \cite{liu2010multiple,ekrem2011secrecy}.  
 Let 
$\mathbf{H}_1 \in \mathbb{R}^{n_1 \times n_t}$ and $\mathbf{H}_2 
\in \mathbb{R}^{n_2 \times n_t}$ be the channel matrices corresponding to  
user~$1$ and 
user~$2$.  Then, the received signals at user~$1$ and user~$2$, respectively, can be represented as
\begin{subequations}\label{eq:signal model}
	\begin{align} 
	\mathbf{y}_1  &= \mathbf{H}_1(\mathbf{x}_1 +  \mathbf{x}_2) +\mathbf{z}_1,\\
	\mathbf{y}_2 &= \mathbf{H}_2(\mathbf{x}_1  + \mathbf{x}_2) + \mathbf{z}_2,
	\end{align}
\end{subequations}
in which $\mathbf{z}_1 \in \mathbb{R}^{n_1 \times 1}$ and 
$\mathbf{z}_2 \in \mathbb{R}^{n_2 \times 1}$ are two independent identically 
distributed (i.i.d) Gaussian noise vectors  with mean zero and identity covariance matrices.

\subsection{Secrecy Capacity Region}
The secrecy capacity region of this channel under a matrix constraint on the covariance matrix of the input $ \mathbf{x}$ is proved in \cite[Theorem~1]{liu2010multiple}. 
However, in practical MIMO systems, a  total power constraint  $P$ at the transmitter is more common. Under this assumption, using Corollary~1 in \cite{ekrem2012capacity}, the  \textit{secrecy capacity region} of this channel can 
be represented as  
\begin{subequations} \label{eq: mathmodel}
	\begin{align}
	R_1 &\leq  \frac{1}{2}\log|\mathbf{I}_1 + {\mathbf{H}_1 \mathbf{Q}_1 \mathbf{H}_1^T}|-\frac{1}{2}\log|\mathbf{I}_2 +  \mathbf{H}_2 \mathbf{Q}_1 \mathbf{H}_2^T|, \label{eq: mathmodel_confi1} \\
	R_2 &\leq \frac{1}{2} \log \bigg|\mathbf{I}_2 + \frac{\mathbf{H}_2 \mathbf{Q}_2 \mathbf{H}_2^T}{\mathbf{I}_2 +\mathbf{H}_2 \mathbf{Q}_1 \mathbf{H}_2^T}\bigg| \nonumber \\ 
	& \qquad\qquad- \frac{1}{2}\log \bigg|\mathbf{I}_1 + \frac{\mathbf{H}_1 \mathbf{Q}_2 \mathbf{H}_1^T}{\mathbf{I}_1 +\mathbf{H}_1 \mathbf{Q}_1 \mathbf{H}_1^T}\bigg|,  \label{eq: mathmodel_confi2} \\
	&\textmd{s.t.} \quad {\rm tr}(\mathbf{Q}_1)+ {\rm tr}(\mathbf{Q}_2) \leq P,\; \mathbf{Q}_1\succcurlyeq \mathbf{0},\;\mathbf{Q}_2\succcurlyeq \mathbf{0},\label{eq: mathmodel_const}
	\end{align}
\end{subequations}
in which 
$R_i$, $i \in\{1,2\}$ is the secure achievable rate at user~$i$,   $\mathbf{I}_i$ is an identity matrix of size $n_i$, and $\mathbf{Q}_i$ is the 
covariance matrix of $\mathbf{x}_i$. 
%
By definition,  $\mathbf{Q}_i = \mathbb{E}(\mathbf{x}_i\mathbf{x}_i^{T})$ where $\mathbb{E}(\cdot)$ denotes expectation, and thus, $\mathbf{Q}_i$ is positive semidefinite, i.e., $\mathbf{Q}_i \succcurlyeq \mathbf{0}$.
Further, since total transmit power cannot be higher than  $P$, we have  $P \ge \mathbb{E}(\mathbf{x}\mathbf{x}^{T}) = {\rm tr}(\mathbf{Q}_1)+{\rm tr}(\mathbf{Q}_2).$

\subsection{Existing Solutions}
Although the secure capacity region of the two-user MIMO-NOMA channel is given in \eqref{eq: mathmodel}, it is still unknown how to analytically form  $\mathbf{Q}_1$ and $\mathbf{Q}_2$ to achieve the capacity region. This is because the right-hand side expressions both in \eqref{eq: mathmodel_confi1} and \eqref{eq: mathmodel_confi2} are non-convex, and thus, the corresponding optimization problems are challenging.     
Early works like \cite{liu2010multiple}  use an exhaustive search over all possible $\mathbf{Q}_1$ and 
$\mathbf{Q}_2$, satisfying  the constraints in \eqref{eq: mathmodel_const}. Such an approach is, however, prohibitively complex for practical systems. Later, Fakoorian \textit{et al.} \cite{fakoorian2013optimality} proposed    
a  GSVD-based precoder for this problem. The rate region of this method is, however, far from the capacity region when $n_1$ and/or $n_2$ are small numbers and, in particular, when the users have a single antenna. Intriguingly, such cases are very prevalent and thus important in practice.  

Lately,  another  approach was proposed to solve this problem  \cite{qi2020secure} whose achievable rate region is very close to the optimal solution for any number of antennas at each node.
The main observation is that the two-users MIMO-NOMA channel can be decomposed into two MIMO \textit{wiretap channels} by splitting power between the two users. Then, the associated optimization problems are solved one at a time. Despite the fact that the rate region obtained from this solution is very close to the secure capacity region of the channel, this solution incurs an unacceptable amount of delay, which hinders it from being used for practical systems. To be specific, finding optimal $\mathbf{Q}_1$ and $\mathbf{Q}_2$ could take 
several hundred milliseconds, \cite{qi2020secure} wheres the \textit{coherence time} of the wireless channel can  be as small as a few milliseconds \cite{tse2005fundamentals}. 
That is, the solution assumes that the channel is constant for several hundred milliseconds while it changes much faster in practice.

In this letter, we propose a DL-based signaling design to approach the secure capacity of the MIMO-NOMA channel within a practically acceptable delay. This is obtained at the expense of a slightly smaller achievable rate region. In the following, we describe the structure of the DNN, the training process, and the test results.

\section{Deep Learning-based Solution}\label{sec:deep}

In this section, we build a supervised deep learning model to determine suitable covariance matrices for  \eqref{eq: mathmodel}.
Put differently, we describe signaling design (precoding  and power allocation) for secure transmission over the MIMO-NOMA channel. We present a DNN structure that includes
data generation, training methodology, DNN structure, and hyper-parameters.

\subsection{Data Generation and Labeling}
In  \textit{supervised learning}, labeled data is used for drawing inferences, i.e., for classification,  regression, or approximation. 
In this paper, labeled training data is used to build predictive models to learn the functions mapping inputs to outputs, i.e., to approximate optimal signaling in the secure MIMO-NOMA by regression. Specifically, for each set of $\mathbf{H}_1$ and $\mathbf{H}_2$ we  find $\mathbf{Q}_1$ and $\mathbf{Q}_2$
and use them for training a DNN and determining a model.

We  decompose the secure MIMO-NOMA channel into two MIMO wiretap channels  \cite{qi2020secure}.
For $\alpha \in [0, 1]$, we allot $\alpha P$ and $(1-\alpha) P$ to  user~$1$  and  user~$2$, respectively. Then, we find the 
covariance matrix $\mathbf{Q}_1$ from \eqref{eq: mathmodel_confi1}, i.e., 
\begin{subequations}\label{eq:R1}
	\begin{align}
	\mathbf{Q}_1^* &= \arg \max \limits_{\mathbf{Q}_1} 
	\frac{1}{2}\log \frac{| \mathbf{I}_1 + {\mathbf{H}_1 \mathbf{Q}_1 
			\mathbf{H}_1^T}|}{|\mathbf{I}_2 +  
		\mathbf{H}_2 \mathbf{Q}_1 \mathbf{H}_2^T|}, \label{eq:R1C1} \\
	&{\rm s.t.}\quad \mathbf{Q}_1 \succeq 
	\mathbf{0}, \;  {\rm tr}(\mathbf{Q}_1)\leq P_1 = \alpha P. \label{eq:R1C2}
	\end{align}
\end{subequations}
Now, this problem can be seen as a wiretap channel in which user~$1$ is the legitimate user and user~$2$ is an eavesdropper.  Thus, we can apply any wiretap channel solutions to solve it. Alternating optimization and
water filling (AOWF) algorithm \cite{li2013transmit} and rotation-based \cite{zhang2019rotation}  method are two of them. Once $\mathbf{Q}_1^*$ is obtained, we plug it  in  \eqref{eq: mathmodel_confi2} and manipulate it to get 
\begin{subequations} \label{eq:R2new}
	\begin{align}
	\mathbf{Q}_2^* &=\arg \max \limits_{\mathbf{Q}_2}  \frac{1}{2}\log 
	\frac{|\mathbf{I}_2 +  \mathbf{H}^{\prime}_2 \mathbf{Q}_2 
		\mathbf{H}^{\prime T}_2|}{
		|\mathbf{I}_1 +  \mathbf{H}^{\prime}_1 \mathbf{Q}_2 
		\mathbf{H}^{\prime T}_1|}, \label{eq:R2C1}\\
	&{\rm s.t.}\quad \mathbf{Q}_2 \succeq \mathbf{0}, \; {\rm tr}(\mathbf{Q}_2)\leq P_2 = (1-\alpha) P,  \label{eq:R2C2}
	\end{align}
\end{subequations} 
in which $\mathbf{H}^{\prime}_i \triangleq 
\mathbf{\Lambda}^{-\frac{1}{2}}_i\mathbf{V}^{T}_i\mathbf{H}_i$, $i\in\{1,2\}$, where $\mathbf{\Lambda}_i$ and $\mathbf{V}_i$	are obtained from eigenvalue decomposition of $\mathbf{I}+\mathbf{H}_i
\mathbf{Q}^{*}_1
\mathbf{H}_i^T$, i.e., $\mathbf{I}+\mathbf{H}_i
\mathbf{Q}^{*}_1
\mathbf{H}_i^T=\mathbf{V}_i\mathbf{\Lambda}_i\mathbf{V}_i^T$. 
Then again, \eqref{eq:R2new}  is the rate for a MIMO wiretap channel where  $\mathbf{H}^{\prime}_2 $ and $\mathbf{H}^{\prime}_1 $ are the channels corresponding to the legitimate user and eavesdropper, respectively. Thus, we solve it using a wiretap solution. {Although this approach is suboptimal, the resulting rate region is close to the optimal solution--obtained by a brute-force search.} 

We next describe the structure of the DNN used for finding suitable covariance matrices for the MIMO-NOMA networks.


\subsection{Network Structure} 
\begin{figure}[h]
	\centering
	\includegraphics[width=0.48\textwidth]{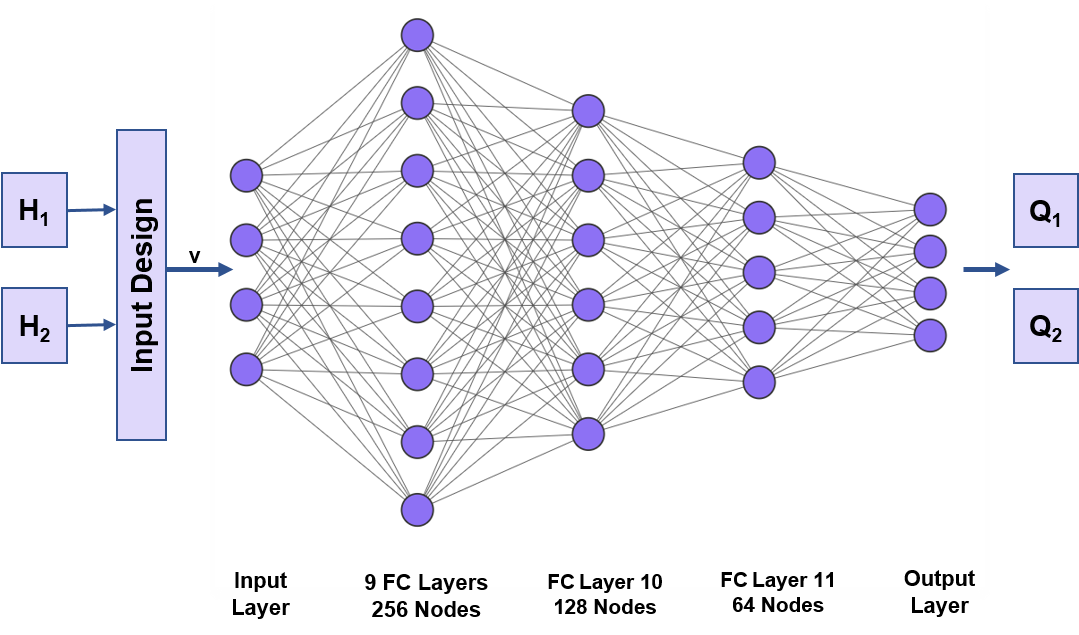}
	\caption{{The  structure of the multi-layer perceptrons network used.}}
	\label{fig:net}
\end{figure}

{We use  a \textit{multi-layer perceptrons} (MLP) DNN in this paper. As  feed-forward neural nets, MLPs are less complex,  easy to design, and have quick run time.} The  structure of the network is shown in Fig.~\ref{fig:net}. As we will see in Section~\ref{sec:pre}, the input is a feature mapping of  $\mathbf{H}_1$ and $\mathbf{H}_2$--the channel matrices of user~1 and  user~2.
\textit{Rectified linear unit} (ReLU) \cite{xu2015empirical}
serves as the activation function.
{
	ReLUs are sparse and have a reduced likelihood of vanishing gradient which reduce training and inference time for neural networks.}  The network has nine fully  connected (FC) layers, each with a width of 256 nodes. The network then funnels through a 128 node layer  and a 64 node layer before reaching the output layer. The output layer is the upper triangular elements of the covariance matrices ($\mathbf{Q}_1$ and $\mathbf{Q}_2$) that the network is trying to learn how to predict. We note that since $\mathbf{Q}_1$ and $\mathbf{Q}_2$ are symmetric, once we get the
upper triangular elements, we know all elements. The size of the output layer depends on $n_t$ and is equal to $n_t(n_t+1)$. 


We have investigated the effect of various hyperparameters including learning rate, drop factor, Adam optimizer, validation frequency, mini-batch size, and validation patience  in order to most effectively train the network. The final, tuned hyperparameters are shown in Table~\ref{table:tab_param}.

\begin{table}[h]
	\caption{Hyper-parameters.}
	\label{table:tab_param}
	\centering
	\begin{tabular}{|l|c||l|c|}
		\hline
		Hyper-parameter        & Value  & Hyper-parameter & Value \\ \hline
		Initial learning rate  & 0.001  & Mini batch size        & 256  \\
		Learn rate {drop} factor & 0.5 & Learn rate {drop} period & 5  \\
		Training set size & $5 \times 10^5$ & Validation set size & 
		$10^5$\\
		Validation frequency & $10^3$ iters& Validation patience & 5\\
		\hline
	\end{tabular}
\end{table}

\subsection{Pre-processing}\label{sec:pre}

\subsubsection{Input design} A big advantage of DL algorithms is that they reduce/eliminate the need for feature engineering as they try to learn high-level features from data. Hence, this problem the input could simply be the channel matrices  $\mathbf{H}_1$ and $\mathbf{H}_2$. However, we have observed that some nonlinear combinations of the channels improve the network performance. Specifically, observing that $\left|
\mathbf{I}_{i}+\mathbf{H}_i\mathbf{Q}_i\mathbf{H}_i^T
\right|=\left|
\mathbf{I}_{i}+\mathbf{H}_i^T\mathbf{H}_i\mathbf{Q}_i
\right|$, we can rewrite \eqref{eq:R1} and \eqref{eq:R2new} as functions of $\mathbf{H}_i^T\mathbf{H}_i$. Then, the input is
designed based on $\mathbf{H}_i^T\mathbf{H}_i$, not  $\mathbf{H}_i$  \cite{zhang2019deepwiretap}.  This makes our design independent of the number of antennas {at the} users since the size of  $\mathbf{H}_i^T\mathbf{H}_i$ is  $n_t \times n_t$ which does not depend on $n_1$ and $n_2$, unlike the size of $\mathbf{H}_i$. This also simplifies the inputs. 

\subsubsection{Scaling} Before feeding the data to the network, we scale it to avoid over-fitting and improve the performance \cite{bishop1995neural}.
It makes back-propagation more efficient \cite{lecun2012efficient}, and allows the network to more quickly learn the optimal parameters for each input node.
Normalizing or standardizing the inputs are the two common ways of scaling.
We normalize the input variables.  To summarize, the input vector $\mathbf{v}$ is 
designed as
\begin{align}\label{eq:inputVec}
\mathbf{v} = [0.05\mathbf{v}_1, 
0.002\mathbf{v}_2]^T,
\end{align}
in which $\mathbf{v}_1$  and  $\mathbf{v}_2$ are 
given by
{\begin{subequations}
	\begin{align}
	&\mathbf{v}_1 =        {\rm vec}([\mathbf{H}_1^T\mathbf{H}_1 \;\;
	\mathbf{H}_2^T\mathbf{H}_2]), \label{eq:inputV1}\\
	&\mathbf{v}_2 =       {\rm vec}([\mathbf{H}_1^T\mathbf{H}_1 \;\;
	\mathbf{H}_2^T\mathbf{H}_2]^T[\mathbf{H}_1^T\mathbf{H}_1 \;\;
	\mathbf{H}_2^T\mathbf{H}_2]), \label{eq:inputV2}
	\end{align}
\end{subequations}
\noindent where $\rm vec(\mathbf{A})$ converts matrix  $\mathbf{A}$ to a vector.
The coefficients of $\mathbf{v}_1$ and  $\mathbf{v}_2$ in \eqref{eq:inputVec}  are chosen based on the histogram of all elements of $\mathbf{v}_1$ and  $\mathbf{v}_2$ for $10^3$ input channels.
After this normalization,   with a high probability,  the elements of  $\mathbf{v}$ will be in the range of $[-1\;\;1]$.
} 

\section{Numerical Results}\label{sec:decomp}

We next evaluate the performance of the proposed DNN-based covariance matrix design (precoding and power allocation) in
different antenna settings. A large dataset was used to have the network generalize for many different channels, and cross-validation data was used to prevent over-fitting. The test dataset was based on 1,000 channel matrices $\mathbf{H}_1$ and 
$\mathbf{H}_2$ whose elements were generated randomly based on $\mathcal{N}(0,1)$.

The performance of the proposed solution can be evaluated in different ways. We may find the mean square error (MSE) of elements $\mathbf{Q}_1$ and $\mathbf{Q}_2$ provided by the DNN and those obtained from a traditional capacity-approaching. The lower the MSEs, the better the regression. Alternatively, we can substitute $\mathbf{Q}_1$ and $\mathbf{Q}_2$ to \eqref{eq: mathmodel} and compare the secure rate region achieved by the DNN with analytical methods like GSVD and maximum achievable rates (capacity) for this channel. We also compare the performance of the DL and traditional iterative methods in terms of computation time.

\begin{figure}[t]
	\centering
	\begin{minipage}[t]{0.39\textwidth}
		\centering
		\includegraphics[width=\textwidth]{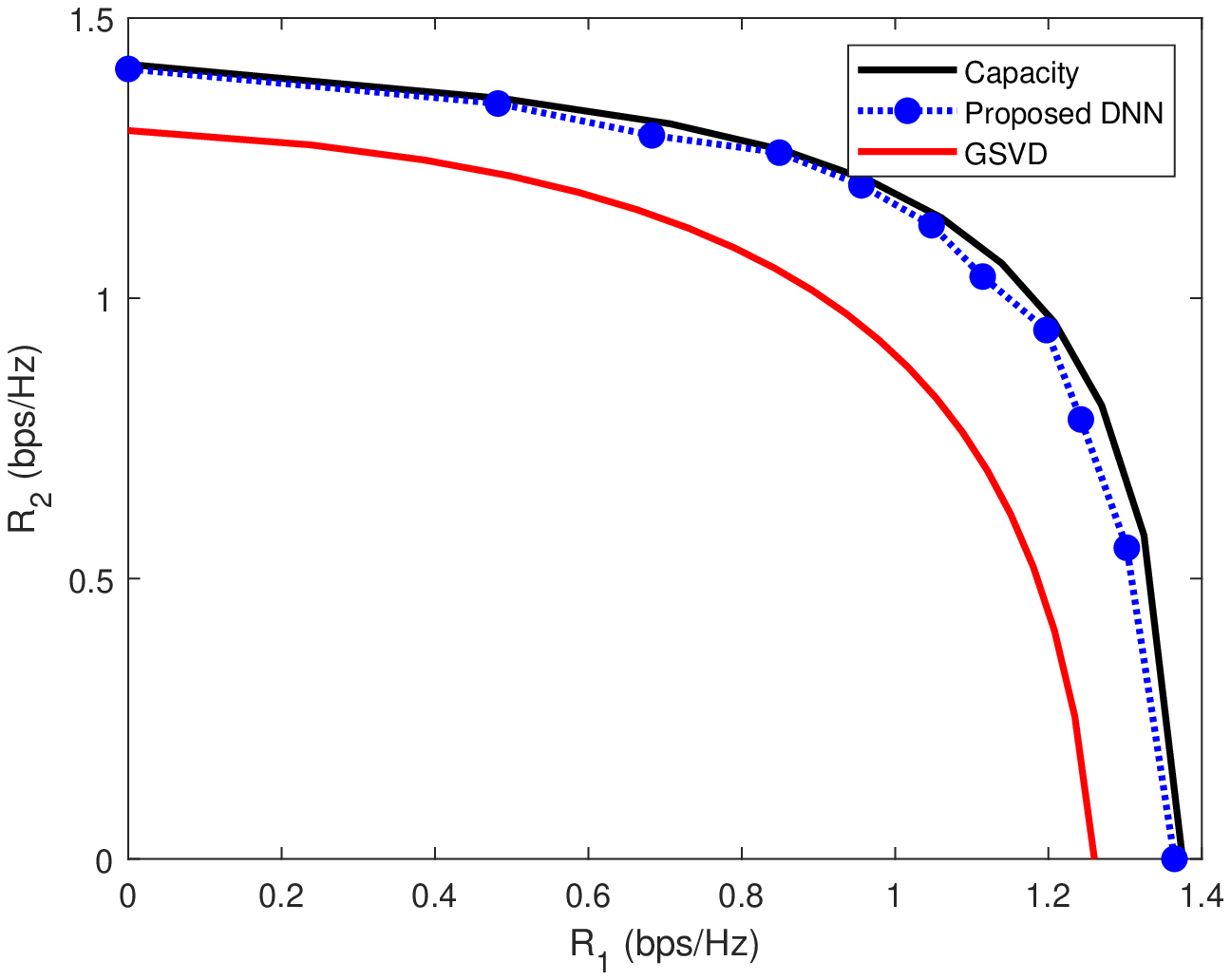}
		\caption{Secrecy rate regions for 
			$n_t=2, n_1=n_2=1.$ }
		\label{fig:twotx_rateregion}
	\end{minipage} 
	\begin{minipage}[t]{0.39\textwidth}
		\centering
		\includegraphics[width=\textwidth]{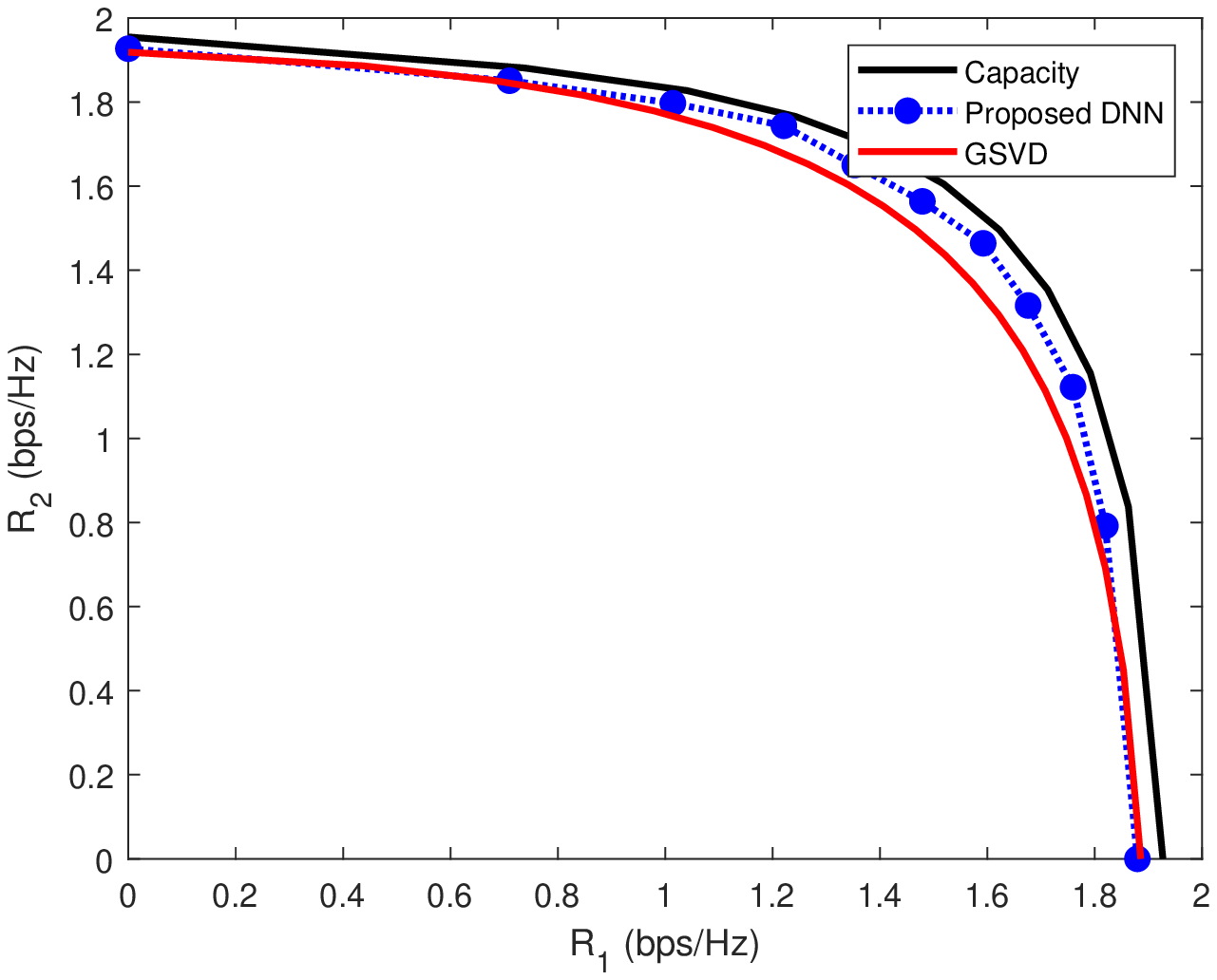}
		\caption{Secrecy rate regions for 
			$n_t=3, n_1=n_2=1.$  }
		\label{fig:threetx_rateregion}
	\end{minipage} 
\end{figure}

Figures~\ref{fig:twotx_rateregion} and \ref{fig:threetx_rateregion} illustrate the accuracy of the proposed DNN model compared to GSVD and capacity region. The rate region plots were made for $n_t=2$ and $n_t=3$. The users have a single antenna, and the BS power is 10 Watts.  Our DNN is able to find covariance matrices for any given $\alpha \in [0,1]$. To demonstrate this, rate region plots were made. These plots show the different rates simultaneously achieved by user~1 and user~2 as the power splitting factor $\alpha$ is changed. 
In order to be accurate at different power splitting factors, eleven networks were trained, each at a different alpha starting from $\alpha=0$ with a step size of 0.1. This produces a smooth, piecewise linear secure rate region curve and proves that the neural network can be generalized for any $\alpha$.
It is important to note that different values of $\alpha$
correspond to different services and result in very different covariance matrices. 
 For example, $\alpha=0$ implies  $\mathbf{Q}_1=\mathbf{0}$ which gives $R_1=0$, i.e., security is important only for user~2, whereas $\alpha=1$ has the reverse implication. That being said, we choose the value of $\alpha$ and the associated DNN based on the users' quality of service. 
  {Averaging over all $\alpha$s, the DNN achieves  \%98.9
  	and \%97.7 of the capacity rates for  $n_t = 2$ and $n_t = 3$, respectively. } We have used the same regularization parameters on all networks. A finer tune would increase this accuracy. 

As can be seen in Fig.~\ref{fig:twotx_rateregion} and \ref{fig:threetx_rateregion}, the DNN highly outperforms analytical methods like GSVD in terms of rate region.
Further, the proposed method largely outperforms existing iterative solutions \cite{park2015weighted,qi2020secure} in terms of computation time. 
Table \ref{table:time_results} shows the input signaling design time gain that we obtain with the DNN. It is several times faster than the rotation algorithm and AOWF. 
All algorithms are tested on the same machine in {\sc Matlab}.
The time difference becomes much larger as $n_t$ increases. In short, the DNN is able to get very close to the capacity region of the MIMO-NOMA  and achieves this much faster, and thus, it is a viable solution. 

\begin{table}[]
	\caption{Solution time in milliseconds}
	\centering
	\label{table:time_results}
	\begin{tabular}{|l|l|l|l|l|}
		\hline
		$n_t$ & \Tstrut Rotation & AOWF    & DNN      \\ \hline\hline
		2         & 6.6 & 22.7 & 3.7 \\ \hline
		3         & 26.1 & 31.4 & 6.3  \\ \hline
	\end{tabular}
\vspace{-.1cm}
\end{table}

\section{Conclusion}\label{sec:con}

A novel deep learning assisted covariance matrix design for the two-user MIMO-NOMA with confidential messages has been developed, trained, and tested in this paper. 
The proposed DNN, which is used to approximate
the capacity region of this channel, is able to achieve nearly perfect accuracy for maximum secure rates for this channel. Remarkably, using the DNN for signaling design significantly reduces the solution time versus existing iteratively solutions 
and brings this time low enough that it can be used in practice. It also significantly outperforms GSVD precoding in achievable secure rates.

\typeout{}

\balance

\end{document}